Searching for the "Holy Grail" of sponsorship-linked marketing:

A generalizable sponsorship ROI model

Jonathan A. Jensen, Ph.D.

Corresponding author. Jonathan A. Jensen, Associate Professor, Sport Administration, Department of Exercise and Sport Science, 308 Woollen, University of North Carolina at Chapel Hill, Chapel Hill, NC, 27599-8700, Phone: 919-962-0959, Email: jajensen@email.unc.edu.

Acknowledgments: The research was supported by an IBM Junior Faculty Development Award from the University of North Carolina at Chapel Hill. Thanks go to the many students whose work supported this research, including David Head, Rebecca Lawson, Connor Leeson, Steven Melson, Christopher Mergy, Akash Mishra, Mo Moreau, Olivia Monroe, George Schmitt, Garrett Shifflett, and Jeremy Vlacancich.

*Abstract:*

Marketers routinely allocate a significant portion of their budget to sponsorship. However, isolating the return on investment (ROI) from such efforts has remained a challenge. Thus, a dataset of more than 5,800 sponsorships is used to develop a model that utilizes the sponsor's renewal of the sponsorship as a proxy for positive ROI. In addition to its contribution to the sponsorship-linked marketing literature, the resulting model can be utilized by managers to generate predicted values relative to the sponsor's probability of renewal and the ultimate duration of time the sponsor will remain with the property, representing a novel managerial contribution.

*Management Slant:*

- Sponsorship ROI has been identified by marketing practitioners as the "Holy Grail" of non-traditional marketing measurement
- This study's model uses the sponsor's decision to renew the sponsorship as a proxy for positive ROI, ensuring generalizability to every type of sponsor and property
- The employment of predictive analytics to sponsorship sales and strategy allows for the model's use as a revenue projection tool for sport properties
- Managers can use to model to perform sponsorship portfolio audits, and allocate resources towards sponsors most likely to exit in the next 1-2 years

Searching for the "Holy Grail" of sponsorship-linked marketing:
A generalizable sponsorship ROI model

For decades, the development of a model assessing which factors are predictive of positive return on investment (ROI) for the brands that invest more than $65 billion annually in sponsorship has long been compared to the quest for the "Holy Grail" (Bednar, 2005; IEG, 2018). "Measurement is the holy grail of all marketing activity," wrote Caroline Campbell of Australian-based Event Sponsors (2021). "What is a little more unknown is what the return on investment (ROI) will be prior to activating a sponsorship campaign. This insight would most certainly help a potential sponsor in the decision-making process." Numerous professionals in the field have suggested that an answer may lie in predictive analytics. "One thing that is certain is predictive analytics is a core component to identifying holistic sponsorship ROI," noted Lyndon Campbell, Senior Vice President of leading data analytics firm MarketCast (MarketCast, 2021).

Among the oft-cited and inherent challenges of such efforts are differing measurement objectives across the thousands of different firms around the globe investing in sponsorship, as well as the thousands of different types of sponsorships and sponsored properties. In addition, most internal metrics are not only customized to meet each brand's objectives, but also proprietary. Thus, achieving one ROI metric that can be applied, or is generalizable, across all types of sponsors and sponsorships has remained elusive (Jensen & Cobbs, 2014). "Judging from my conversations with sponsors and properties, I haven't found anybody who is truly satisfied with their ROI measurement on an individual sponsorship persistently across time," wrote Steve Seiferheld of Turnkey Intelligence (2010).

This is not uncommon, as generalizability is a consistent issue within the scientific process, across a multitude of different academic fields. This challenge is particularly acute throughout the sponsorship-linked marketing literature, as findings specific to some sponsorship contexts may not be generalizable to others, and results that may be applicable to firms located in one particular country may not be generalizable to those in other regions of the world. For example, one prior solution that has been offered involved investigating the effects of sponsorship announcements on the sponsoring firm's stock price, utilizing the market's reaction to the announcement of the partnership as a proxy for positive returns. The approach revealed meaningful insights across a variety of different types of sponsorships, including those of the Olympic Games (Miyazaki & Morgan, 2001), the Paralympics (Ozturk & Kocak, 2004), teams participating in the Indianapolis 500 (Cornwell et al., 2001), NASCAR teams (Pruitt et al., 2004), naming rights sponsorships (Clark et al., 2002), league sponsorships (Cornwell et al., 2005), title sponsorships (Clark et al., 2009), F1 Racing teams (Cobbs et al., 2012), cycling teams (Drivdal et al., 2018), international football matches (Bouchet et al., 2015), and teams (Bouchet et al., 2017). However, these types of studies are oftentimes challenged by small sample sizes, including analyses of the effects of just 16 (Ozturk & Kocak, 2004), 24 (Pruitt et al., 2004), 25 (Eryigit & Eryigit, 2019), 39 (Becker-Olsen, 2003), 46 (Drivdal et al., 2018), and 49 (Clark et al., 2002) sponsorship announcements in total. Further, a number of these studies have analyzed sponsorships of properties based in specific geographic areas, such as an analysis of sponsorships of four sport clubs in Turkey (Eryigit & Eryigit, 2019) and results specific to 19 firms engaging in sponsorship in South Africa (Kruger et al., 2014). Thus, it is difficult to assume that findings emanating from studies in specific markets such as Turkey and South Africa would be generalizable across other countries in the Middle East or Africa, and beyond to Asia, Europe, and North America.

      Attempts to generalize findings from these studies are challenging, given that the results are not only specific to one type of sponsorship but also not generalizable beyond firms that are publicly-traded corporations. In an attempt to create a generalizable model, Mazodier and Rezaee

(2013) constructed an initial dataset of more than 1,642 sponsorship announcements involving a wide variety of different types of sponsorships involving sponsors from a variety of geographic regions. However, the dataset ultimately was reduced down to a final sample of just 293 sponsorships, once announcements involving privately-held firms and those whose announcement date could not be verified were omitted. This represented less than 20% of the sponsorships in the original sample. It was not until a meta-analysis was employed, analyzing results from 34 such studies spanning more than 20 years, that it became evident that positive returns were not realized in the post-announcement window (Kwon & Cornwell, 2020). This result suggests that much of the findings from prior research were context-specific, and lacked the generalizability necessary to be applied to all sponsorships.

To overcome these challenges, this study seeks to generate insights relative to determinants of ROI from sponsorship that are generalizable across every major type of sponsorship, geographic region, and sponsor category. As a proxy for positive ROI and to ensure the model can be applied to any sponsor and sponsorship, we utilize the sponsor's decision to either renew or exit a sponsorship as the dependent variable in the model, with the underlying assumption that, with few exceptions, sponsoring firms would not continue to renew sponsorships using scarce marketing resources that did not generate positive ROI. There is a strong foundation in the literature for the link between a sponsor's decision to continue a sponsorship and positive effects for sponsors. Cornwell, Roy, and Steinard (2001) explained that the longer a sponsorship continues, the greater the potential for the sponsorship to become a source of competitive advantage, based on its ability to influence consumer-based outcomes. In another example, research has suggested that the longer a sponsorship continues, the greater the association between the sponsoring brand and property in a consumer's memory (Cornwell & Humphreys, 2013; Johar & Pham, 1999). "Seeing a sponsor's name associated with the same sporting event, year after year, gives the consumer multiple opportunities to elaborate about the significance of the product-sponsorship relationship, thus creating stronger associations in memory," explained Cornwell et al. (2001, p. 42).

In the development of a generalizable model, we constructed a historical dataset featuring a pooled sample of more than 5,800 different sponsorships (spanning more than 23,000 total observations). Included within the dataset are global sponsorships of the Olympics and World Cup, sponsorships of teams in Formula One (F1), shirt sponsorships of teams in the Bundesliga, English Premier League (EPL), La Liga, Ligue 1, and Serie A, Major League Soccer (MLS) and the National Basketball Association (NBA), official status (i.e., league) sponsorships of Major League Baseball (MLB), Major League Soccer (MLS), the National Association for Stock Car Auto Racing (NASCAR), the NBA and the Women's National Basketball Association (WNBA), the National Football League (NFL), the National Hockey League (NHL), title sponsorships across college football, golf, and tennis (PGA, LPGA, ATP, and WTA), and naming rights sponsorships of venues hosting teams in MLB, MLS, NBA, NFL, NHL, the Bundesliga, EPL, La Liga, Ligue 1, and Serie A (Table 1). Importantly, the complete history of all such sponsorships were compiled, which ensures sampling bias would not be an issue. Not only does this represent the largest dataset in the sponsorship literature, more than five times as large as the 1,077 sponsorships analyzed by Cobbs et al. (2017), but heterogeneity in the types of sponsorships, sponsoring firms, properties sponsored, sponsor categories, and the geographic regions represented helps to ensure that findings are generalizable across all sponsorships. Given the potential for the effects of the COVID-19 pandemic to influence decision-making, the data used to train the model ends at the start of the 2020 calendar year. This was prior to the effects of the start of the pandemic in March 2020, ensuring that any decision-making was not pandemic-influenced.

## Theoretical Background

Prior research would suggest that regional proximity between the sponsoring brand and sponsored property, such as the instance of a firm's corporate headquarters being located in the same city as the sponsored organization, should lead to greater ROI (Woisetschläger et al., 2017). Theoretically, sponsorships with regional proximity signal to consumers a high level of commitment from both sides to the relationship. In the specific context of "hometown" naming rights agreements, Clark et al. (2002) noted that regional proximity may also allow the sponsor to more effectively activate the sponsorship. Both Jensen and Cornwell (2018) and Jensen et al. (2020) found regional proximity was a statistically significant predictor of sponsorship renewal. Woisetschläger et al. (2017) suggested that a higher level of commitment on behalf of a hometown sponsor may affect consumer perceptions of fit and influence whether sponsor motives are seen as affective, rather than calculative, leading to positive ROI.

Congruence, defined as the perceived fit between a brand and property, is one of the most frequently studied theoretical constructs in the sponsorship literature (Fleck & Quester, 2007). The effects of congruence are undergirded by balance theory (Heider, 1958), which suggests that incongruent information is more likely to be ignored by consumers. Consistent with balance theory, research has consistently demonstrated that the better consumers perceive the fit to be between a sponsoring brand and sponsored property, the more likely the sponsor will achieve their intended cognitive, affective, and behavioral effects (Cornwell, Weeks, & Roy, 2005). Thus, it is not surprising that Jensen and Cornwell (2017) and Jensen et al. (2022) found that congruent sponsors were significantly less likely to exit global event and league sponsorships. Research has also suggested that sponsorships which are incongruent may lead to psychological tension in the consumer's mind, and that congruence can lead to more positive attitudes towards sponsors (Woisetschläger et al., 2017). Thus, congruence should be suggestive of positive sponsorship

ROI.

Brand equity, defined by Keller (1993) as the potential effect of brand knowledge on a consumer's purchase decision, is a frequently noted objective of sponsorship-linked marketing activities. Research has proven that managers engaging in sponsorship believe that longer-term sponsorship relationships should improve the potential that the investment will increase the sponsoring brand's equity (Cornwell, Roy, & Steinard, 2001). Jensen and Cornwell (2017) and Jensen et al. (2020) found high levels of brand equity significantly reduced the probability of sponsorships ending. Thus, it is expected that sponsorships involving brands with a higher degree of brand equity not only may be more patient in their efforts designed to nurture the firm's brand, leading to longer-term engagements, but should lead to improved ROI.

A firm's business-to-business (B2B) perspective often leads it to take a longer-term perspective. In addition, prior research has found that a sponsoring firm's B2B perspective may affect its returns from sponsorship (Mahar, Paul, & Stone, 2005), its choice of properties to sponsor (Cunningham, Cornwell, & Coote, 2009), and interest in generating sales from sponsorships (Cobbs & Hylton, 2012). In the context of U.S.-based event sponsorships and league sponsorships, Jensen and Cornwell (2018) and Jensen et al. (2022) found B2B firms to be more likely to extend such sponsorships. Thus, a B2B perspective should be suggestive of improved ROI.

Research has established that increased clutter will result in a negative impact on a consumer's ability to recall those sponsors (Breuer & Rumpf, 2012). Cornwell, Relyea, Irwin, and Maignan (2000) found that an increase in perceptions of clutter by consumers had a negative affect on the number of sponsors recognized and recalled. It is also a possibility that an increase in the number of sponsors may result in issues related to servicing each individual sponsor (O'Reilly, Heslop, & Nadeau, 2011). Further, both Jensen and Cornwell (2017) and Jensen et al.

(2022) found clutter detrimental to the renewal of sponsorships. This foundation in the literature suggests that an increase in clutter should result in decreased ROI.

The model also controls for a wide variety of different potentially confounding variables, including economic conditions in the home country of the sponsor (e.g., Jensen & Cornwell, 2017), the location of the firm's corporate headquarters (either Africa, Asia, Australia, Europe, North America, or South America), and whether the firm is publicly traded or privately owned (Miller et al., 2011; See Table 2 for descriptive statistics). In addition to the presence of high technology (e.g., Clark et al., 2002), other product categories for which high demand for sponsorship exists (i.e., alcoholic and non-alcoholic beverages, automotive, banks, insurance, etc.) was also controlled for in the model by inserting variables indicating what IEG (2016) determined to be the categories most active in sponsorship. The model also controls for the presence of high demand properties, such as global mega-events like the Olympic Games and World Cup, as well as the "Big Four" North American sport leagues (MLB, NBA, NFL, and NHL; Jang et al., 2020).

## Method

The methodology utilized is survival analysis modeling, traditionally utilized in academic fields such as public health and biostatistics to test the potential influence of covariates on a person's lifetime (Box-Steffensmeier & Jones, 2004). Alternatively known as time-to-event regression modeling, survival analysis is also referred to as event history analysis (sociology), duration analysis (econometrics), and failure-time analysis (engineering), and allows for the isolation of both whether an event in question has taken place as well as the duration of time until such event occurs (Box-Steffensmeier & Jones, 2004). The novel approach employed in this use case is to utilize as the dependent variable in the model the probability of the sponsor exiting the sponsorship, in each and every year, as a proxy for a continued positive ROI. In other academic fields such as political science, survival analysis has previously been applied to analyze time-to-event duration data ranging from United Nations peacekeeping missions, military interventions, and the careers of members of Congress (Box-Steffensmeier & Jones 2004). Survival analysis is particularly effective for the analysis of longitudinal data, for two main reasons. First, such models are robust to the presence of censored observations, or observations for which the final duration is unknown, given that they are currently ongoing. Second, it can analyze the effect of either time-invariant or time-varying covariates (Singer & Willett, 2003). Given this flexibility, the methodology has been previously utilized to investigate the durations of athlete careers in the NBA (Staw & Hoang, 2013; Hoang & Rascher, 1999) and NFL (Volz, 2017), as well as the career tenures of coaches in the NBA (Wangrow et al., 2018), NFL (Salaga & Juravich, 2020), and college basketball (Wangrow et al., 2021). It has also been used to examine the survival of sport organizations (Cobbs et al. 2017) and relationships with donors to intercollegiate athletic departments (Wanless et al., 2019; Jensen, Wanless, et al., 2020).

**Data Analysis Overview**

There are three key concepts that are essential to understanding the concept of survival analysis. The first is the Kaplan-Meier (1958) survivor function estimate ($S(t_{ij})$). Singer and Willett (2003, 334) defined the survivor function as the "probability that individual $i$ will survive past time period $j$." An underlying assumption is that the individual $i$ did not experience the event occurrence of interest in the $j$th time interval, and consequently survives to the end of time period $j$. Said another way, the random variable for time ($T_i$) for individual $i$ exceeds $j$. Accordingly, the survivor function is defined by the formula below:

$$S(t_{ij}) = \Pr[T_i > j]$$

The next concept important to gaining an understanding of survival analysis is the hazard function, or hazard rate. Defined as the rate at which the duration or event ends, given that the target event has not been experienced prior to a particular time internal, the hazard rate is equivalent to the conditional probability of the event occurrence of interest being experienced in each timeframe (Box-Steffensmeier & Jones, 1997). $T_i$ represents the time period $T$ for individual $i$, thus the the discrete-time hazard function is represented by the following formula:

$$h(t_{ij}) = \Pr[T_i = j | T_i \geq j]$$

Defined by Singer and Willett (2003, p. 337) as "that value of $T$ for which the value of the estimated survivor function is .5," the median lifetime is computed using the aforementioned survivor function. The median lifetime is the point at which the survivor function is exactly 0.5, or the point at which half of the individuals in the dataset (in the case of this study, sponsorships) have experienced the event occurrence of interest, and half as of yet have not. In situations in which the survivor function in a particular time period is above or below 0.5 (i.e., when a survivor function of 0.5 falls between two values of $S(t_j)$), a formula provided by Miller (1981) can be utilized to linearly interpolate the exact time period at which the survivor function is 0.5. In Miller's (1981) formula, $m$ represents the last time period in which the survivor function is above 0.5. Further, $\hat{S}(t_m)$ represents the survivor function in that particular time period and $\hat{S}(t_{m+1})$ represents the survivor function for the next time period. The formula is as follows:

$$m + \left[\frac{\hat{S}(t_m) - .5}{\hat{S}(t_m) - \hat{S}(t_{m+1})}\right]((m+1) - m)$$

**Modeling approach**

Given no requirement for an *a priori* parametrization of the baseline hazard function, the Cox proportional hazards model (Cox, 1972) is the most versatile and widely-utilized survival model, and is therefore used here. When working with rather discrete data, which is the case in this study with events occurring once per year, Box-Steffensmeier and Jones (2004) also recommend the Cox model. In addition to coefficients for each covariate, which indicates whether the variable is either increasing or decreasing the probability of event occurrence (in the case of this study, the end of the sponsorship), the Cox model also generates a hazard ratio for each variable. The hazard ratio is interpreted similarly to odds ratios in a logit model with a ratio greater than one reflecting that the variable increases the probability of event occurrence and a value below 1 reflecting a decrease in the probability of event occurrence. Standard errors will be clustered by sponsorship. As indicated below, the Cox (1972) proportional hazards model contains no constant term (β0), which is absorbed into the model's baseline hazard function. Below is an equation that approximates this study's model:

$$h_i(t) = \exp(\beta SponsorshipType_i + \beta EconomicGrowth_{ij} + \beta Inflation_{ij}$$
$$+ \beta SponsorLocation_i + \beta SponsorCategory_i + \beta RegionalProximity_i$$
$$+ \beta Congruence_i + \beta BrandEquity_i + \beta B2B_i + \beta Public_i + \beta Clutter_{ij}) h_0(t)$$

## Results

**Non-parametric results**

We will begin by reviewing non-parametric results, including the dataset's survivor functions, hazard rates, and median lifetime. As described by Singer and Willett (2003), the life table (Table 3) is a helpful way to organize this information, and includes both survivor functions and hazard rates for the data at each time interval, from year one to year 50. To begin, hazard rates in each time period provide a useful glimpse into the proportion of observations that experience the event occurrence of interest in each time period. In the case of this study, the hazard rate equates to the probability of each sponsorship ending in each and every year. As indicated in Table 3, the overall hazard rate across the entire dataset is .2109, equating to a 21.09% probability that a sponsorship will end each year. The inverse of this function is precisely equivalent to a renewal rate, or the probability of a sponsorship being renewed, in each year. Thus, the overall sponsorship renewal rate is 78.9%, or a nearly 79% probability of each sponsor renewing each year.

A review of the results on a per year basis detailed in Table 3 reveal that only in the first three years is the hazard rate larger than the overall rate across the 50-year time period. The hazard rate in year one is .3169 (equating to a 68.3% renewal rate), in year two it is .2531 (a 74.7% renewal rate), and in year three it is .2477, or a 75.2% renewal rate. After the first three years the hazard function continues to steadily decline, from .2074 in year four to .1036 in year seven. Thus, the renewal rate ranges from 79.3% in year four to 89.6% after year seven. A small uptick is observed in year eight (to .1427, or a 85.7% renewal rate), before another steady decline begins after year nine, as the function is reduced from .1121 to .0813 in year 12. The function then remains relatively stable for the next 10 years, ranging between highs of .0817 in year 13 and .0992 in year 20 to lows of .0681 in year 17 and .0602 in year 18, before dropping to only .0241 in year 22. These results represent the fact that only between 6.0-9.9% of all sponsorships lasting past year 10 end over the ensuing decade, equating to a renewal rate ranging from 90.1-93.9% for years 10-21. The hazard rate does not rise again over .10 until year 27, when it reaches .1282 (equating to an 87.2% renewal rate). While it exceeds .10 again in year 30 (.1154), it remains below .10 in every other year through year 40. These results suggest that after the first four years when the hazard function ranges from a high of .3169 to a low of .2074, the probability of a sponsorship ending in each year is steadily reduced to a rate of less than 10% in nearly every year from 10 through year 40. This trend is captured graphically in Graph A of Figure 1, which depicts the hazard function steadily declining from year 10 through approximately year 35.

As stated, survivor functions in each year, defined simply as the percentage of observations that continue for each discrete timeframe, are necessary for computing the median lifetime. The trend over time for the survivor functions included in Table 3 are depicted in Graph B of Figure 1. The graph suggests that there is a significant and precipitous drop in the survivor function over the first 4-6 years, before steadying itself and flattening out between years 7-10. The function then remains fairly flat over the remaining decades. The survivor function at the end of the first year indicates that 68.3% of all sponsorships continue past the first year. The survivor function at the end of the second year, .5102, indicates that just over half of all sponsorships continue past year two. After year three, the function drops to .3838, suggesting that approximately 38% of sponsorships continue past year three.

From an analysis of this data it is evident that the median lifetime, or the point in time at which the survivor function equals .5, is between years 2 and 3. Using the aforementioned formula from Miller (1981) to linearly interpolate the exact point between two intervals finds that the median lifetime is 2.08 years. This result suggests that the time at which half of all sponsorships have ended, and half are still ongoing, is just over two years. Less than 20% of all sponsorship continue past year 7, given that the survivor function is less than .20 after this year (.1901), and dips below 10% after year 14 (.0959). Less than 5% of all sponsorships last past year 23, as the function drops below .05 (.0488) in that time period.

**Semi-parametric results**

We now turn to the results of the model including all of the aforementioned covariates. Five different blocks of variables were entered into a Cox proportional hazards model using a hierarchical procedure. As indicated in Table 4, each group of variables explained a statistically significant amount of incremental variance in the probability of the sponsoring firm exiting the sponsorship, including sponsorship type, $\chi^2(9) = 1299.45$, $p < .001$, economic conditions, $\chi^2(2) = 11.91$, $p = .003$, sponsor location, $\chi^2(5) = 28.60$, $p < .001$, sponsor category, $\chi^2(23) = 105.84$, $p < .001$, and sponsor characteristics, $\chi^2(6) = 125.87$, $p < .001$. An analysis of the Akaike information criterion (AIC) and Bayesian information criterion (BIC) for each of the models at each stage of the model-building process indicates that the final model (Model 5 in Table 3) is the best-fitting model, with an AIC of 76655.3 and a BIC of 77017.9. An analysis of variance inflation factors

(VIFs) for the model's variables finds that the largest VIF is 3.16 and the mean VIF is 1.35, indicating that collinearity is not an issue within the model.

After controlling for differences across the various types of sponsorships within the dataset (Model 1 in Table 4), results indicate that the influence of economic conditions, including both economic growth, $z = 0.37$, $p = .713$, and inflation, $z = 1.88$, $p = .061$, in the home country of each sponsor in every year, are nonsignificant predictors of ROI. Results were generalizable across Australia, Europe, and North America, as significant differences were not found across all three continents. Notably, Asian-based firms are significantly less likely to exit sponsorships, $z = -2.43$, $p = .015$, with the 95% CI excluding 1.0 [.809, .978], suggesting that firms in Asia may be more likely to achieve positive ROI from sponsorship. Results were also generalizable across a wide variety of prominent product categories, with no significant differences found across the alcoholic and non-alcoholic beverage, automotive, tech, hotel, betting, tire, and telecommunications categories. Conversely, significant differences were observed across the airline, $z = -3.19$, $p = .001$, banking, $z = -5.13$, $p < .001$, credit card, $z = -2.00$, $p = .045$, and insurance categories, $z = -4.64$, $p < .001$. Brands in each of these four categories were less likely to exit sponsorships, suggestive of positive ROI across these specific sponsor categories. Conversely, sponsors in the media, $z = 2.71$, $p = .007$, and pharmaceutical categories, $z = 2.07$, $p = .038$, were significantly more likely to exit, suggesting positive results in these categories may be short-lived or that sponsors in these categories are less likely to achieve positive ROI over a longer-term timeframe.

In terms of sponsor-related factors that are statistically significant predictors of positive ROI, results indicate that congruence reduces the probability of a sponsorship ending, suggestive of improved ROI for brands congruent with the sponsored property. Congruent sponsors are 7.5% less likely to end sponsorships, $z = -2.23$, $p = .025$, with the 95% CI excluding 1.0 [.864, .991].

Sponsors enjoying regional proximity with the sponsored property are also less likely to depart, and the effect is statistically significant. Being in the same market as the property decreases the probability a sponsorship will end by 11.6%, $z = -4.08$, $p < .001$, with the 95% CI excluding 1.0 [.833, .938]. As depicted in Graph C in Figure 1, firms with high levels of brand equity are 19.1% less likely to exit sponsorships, suggesting that such firms are more likely to achieve positive ROI. The effect is again statistically significant, $z = -4.38$, $p < .001$, with the 95% CI excluding 1.0 [.735, .889]. Sponsorships with firms exhibiting a B2B perspective are 7.0% less likely to end, $z = -2.25$, $p = .024$, with the 95% CI excluding 1.0 [.873, .991]. Publicly-traded firms are 20.6% less likely to exit, suggesting that large public firms may have additional resources at their disposal to devote to activation, helping to ensure positive ROI. The effect is statistically significant, $z = -8.02$, $p < .001$, with the 95% CI excluding 1.0 [.750, .839]. The influence of clutter is in the expected direction, as every 10 sponsors added increases the probability that each sponsor exits by .13%, perhaps due to the fact that additional clutter reduces a sponsor's ability to generate positive ROI. However, the detrimental effect of clutter falls short of statistical significance, $z = 0.13$, $p = .896$.

## Discussion

Given a recent analysis of the past 20 years of sponsorship research revealed an over-reliance on the consumer effects of sponsorship at the expense of studies that examine the perspective of the sponsoring firm (Cornwell & Kwon, 2020), this investigation of sponsor decision-making importantly helps to fill a gap in the sponsorship-linked marketing literature. The analysis reveals several brand-related characteristics that are predictive of positive ROI, and results that are generalizable across a range of sponsorships and sponsoring firms from a wide variety of different categories and geographic locations. Furthering understanding of the factors that are predictive of the end of sponsorships helps sellers of sponsorship not only target firms that are more likely to enter into longer-term partnerships, but also conditions that may predict the end of relationships with current partners. While the decision to continue or end the sponsorship is only a proxy for positive sponsorship ROI, this research also assists sponsoring firms by illuminating which sponsors may be realizing greater returns from sponsorship, as well as which types of sponsorships and sponsor categories assist firms in realizing a greater degree of success from their investments in sponsorship.

Results revealed several boundary conditions of effects found in prior studies. For example, Clark et al. (2002) hypothesized that tech firms engaged in large scale sponsorships are likely to value the initial signal that the announcement of the partnership provides to potential stakeholders and future shareholders, based on signaling theory. Later, research from Jensen et al. (2020) in the context of naming rights sponsors confirmed this effect, finding that tech firms were more likely to exit such sponsorships given less interest in a longer-term brand-building than the initial signal. However, this study found this effect to not be generalizable beyond that specific type of sponsorship, finding that tech firms are no more or less likely to exit such sponsorships. Another boundary condition identified by this study is that of economic conditions. Jensen and Cornwell (2017) previously found inflation to be a statistically significant predictor of the end of global sponsorships of the Olympic Games. This study found that effects of changes in inflation and economic growth were nonsignificant, and did not influence sponsorships across a much wider span of different types of sponsorships.

**Managerial implications**

Perhaps most importantly, the model produced via this research can be utilized by both sponsoring firms and sponsored properties to make future predictions relative to the propensity of various types of sponsorships to be renewed and generate predicted values that can be used to determine the number of years a sponsor will remain with a property. They may then subsequently utilize the dollar amount allocated towards each sponsorship to forecast the total costs the sponsor may ultimately allocate towards the property, or alternatively the total revenue the property may earn from either an existing or prospective sponsor. Thus, the model can be used by either side of the equation as a financial forecasting tool, for either costs to the sponsor or revenue for the property. With scarce resources available to target potential sponsors, properties can also utilize the model's predictions to project which sponsor categories may ultimately generate the most revenue, based on their predicted ultimate duration as a sponsor. Theoretically, this would allow sales managers to prioritize the highest revenue-producing sponsor categories throughout their proactive sales efforts. Revenue predictions made by the model may also be helpful in determining whether additional salespeople may need to be hired in the future, based on the predicted revenue from the existing sponsor portfolio and when sponsors are predicted to exit the property.

For example, the model predicts that a publicly-traded, North America-based automobile manufacturer with a high degree of brand equity sponsoring the NFL (expecting 2% economic growth and an 6% increase in the consumer price index over the coming year) to ultimately remain with the league for 4.87 years, regardless of the length of the initial agreement. Therefore, if the brand is allocating $2.25 million per year towards the sponsorship, the league would be expected to ultimately earn a total of $10.95 million from the sponsor. A sponsor with the same characteristics in the banking category would be expected to generate $18.77 million in revenue, a difference of nearly $8 million. The same sponsor in the insurance category would generate $16.65 million, while partnering with sponsors in the credit card and retail category would result in revenue of $15.26 and $12.24 million, respectively. Sponsors in the telecommunications and pharmaceutical categories would only generate $10.69 million and $6.82 million, respectively. The millions of dollars in additional revenue resulting from targeting

a sponsor in a different category demonstrates how the application of predictive analytics to the sponsorship sales process can result in significant increases in revenue for sport organizations. In addition, properties can utilize the model to enhance decision-making relative to their proactive sales efforts.

      Managers can also utilize the model to create a sponsorship portfolio audit, generating predictions for the ultimate length of time all current sponsors will stay with the property, as well as how much revenue each current sponsor will ultimately deliver to the sponsored organization, taking into account how much each sponsor is spending and how long each sponsor has already been with the property. The current sponsor portfolio can then by triaged by the probability of renewal, with the model's predictions identifying the subset of sponsors within the portfolio that are most likely to exit their sponsorships within the next year or two, as well as those that are not expected to exit within the next several years. This actionable intelligence can then enhance decision-making relative to the employment of activation managers across the sponsorship portfolio, allowing the most vulnerable sponsors to receive extra attention in their efforts to realize a positive ROI. Alternatively, resources could be re-allocated from sponsors that the model predicts should remain with the property for many years, saving scarce resources that then can be allocated elsewhere. In either scenario, the predictions allow managers to reallocate resources towards sponsor retention efforts, which may ultimately result in increases in sponsor renewal and revenue.

Table 1.
*Overview of dataset*

| Type of Sponsorship | Sponsored Property | n | *N* |
|---|---|---|---|
| The Olympic Partners (TOP) | IOC | 33 | 389 |
| World Cup | FIFA | 47 | 496 |
| Naming Rights | MLB, MLS, NBA, NHL, NFL, Bundesliga, EPL, Ligue 1, La Liga, Serie A | 219 | 2,026 |
| Jersey/Shirt Sponsorships | Bundesliga, EPL, Ligue 1, La Liga, Serie A, MLS, NBA | 774 | 2,474 |
| League Sponsorships | MLB, MLS, NASCAR, NBA, NHL, NFL, WNBA | 551 | 4,156 |
| Event Title Sponsorships | PGA, LPGA, ATP, WTA, College Football | 926 | 5,580 |
| Team Sponsorships | Formula One (F1) | 3,286 | 8,339 |
| Total | | 5,836 | 23,460 |

Table 2.
*Descriptive statistics for dataset*

| Variable | Measure | Count (%) (N = 23,460) | M | SD | Min, Max |
|---|---|---|---|---|---|
| *Sponsorship Type* | | | | | |
| Naming Rights | Binary | 2026 (8.64%) | | | |
| Event Title | Binary | 5580 (23.79%) | | | |
| League | Binary | 4156 (17.72%) | | | |
| Jersey/Shirt | Binary | 2474 (10.55%) | | | |
| Team | Binary | 8339 (35.55%) | | | |
| Olympic Games | Binary | 389 (1.66%) | | | |
| World Cup | Binary | 496 (2.11%) | | | |
| *Economic Conditions* | | | | | |
| Economic Growth (GDP) | Continuous | | 2.27 | 2.08 | -15.02, 34.86 |
| Inflation (CPI) | Continuous | | 3.47 | 37.43 | -22.90, 2947.73 |
| *Sponsor Location* | | | | | |
| Africa | Binary | 30 (0.13%) | | | |
| Asia | Binary | 2582 (11.01%) | | | |
| Australia | Binary | 156 (0.66%) | | | |
| Europe | Binary | 8949 (38.15%) | | | |
| North America | Binary | 11588 (49.39%) | | | |
| South America | Binary | 155 (0.66%) | | | |
| *Sponsor Category* | | | | | |
| Alcoholic Beverage | Binary | 894 (3.81%) | | | |
| Non-Alcoholic Beverage | Binary | 733 (3.12%) | | | |
| Automotive | Binary | 3160 (13.47%) | | | |
| Insurance | Binary | 858 (3.66%) | | | |
| Apparel | Binary | 787 (3.35%) | | | |
| Retail | Binary | 1420 (6.05%) | | | |
| Tech | Binary | 3546 (15.12%) | | | |
| QSR | Binary | 351 (1.50%) | | | |
| Food | Binary | 1027 (4.38%) | | | |
| Media | Binary | 382 (1.63%) | | | |
| Bank | Binary | 1302 (5.55%) | | | |
| Credit Card | Binary | 353 (1.50%) | | | |
| Financial Services | Binary | 845 (3.60%) | | | |
| Medical/Hospitals | Binary | 173 (0.74%) | | | |
| Pharmaceutical | Binary | 100 (0.43%) | | | |
| Personal Care | Binary | 891 (3.80%) | | | |
| Airline | Binary | 569 (2.43%) | | | |
| Shipping/Mail | Binary | 225 (0.96%) | | | |
| Utilities/Power | Binary | 1039 (4.43%) | | | |
| Hotel | Binary | 135 (0.58%) | | | |
| Betting | Binary | 226 (0.96%) | | | |
| Tire | Binary | 571 (2.43%) | | | |
| Telecom | Binary | 1162 (4.95%) | | | |

*Sponsor Characteristics*

| | | | | | |
|---|---|---|---|---|---|
| Regional Proximity | Binary | 7399 (31.54%) | | | |
| Congruence | Binary | 9413 (40.12%) | | | |
| Brand Equity | Binary | 4631 (19.74%) | | | |
| B2B | Binary | 5832 (24.86%) | | | |
| Publicly-Traded | Binary | 15100 (64.36%) | | | |
| Clutter | Continuous | | 13.97 | 14.91 | 1, 59 |

Table 3.
*Life table*

| Period | Time interval | Beginning total | Ended during period | Censored at end of period | Hazard function | Survivor function |
|---|---|---|---|---|---|---|
| 0  | [0, 1)   | 5836 | --- | --- | --- | 1.0000 |
| 1  | [1, 2)   | 5836 | 1849 | 139 | .3169 | .6832 |
| 2  | [2, 3)   | 3848 | 974 | 109 | .2531 | .5102 |
| 3  | [3, 4)   | 2765 | 685 | 98 | .2477 | .3838 |
| 4  | [4, 5)   | 1982 | 411 | 72 | .2074 | .3042 |
| 5  | [5, 6)   | 1499 | 259 | 63 | .1728 | .2517 |
| 6  | [6, 7)   | 1177 | 185 | 37 | .1573 | .2121 |
| 7  | [7, 8)   | 955 | 99 | 36 | .1036 | .1901 |
| 8  | [8, 9)   | 820 | 117 | 43 | .1427 | .1630 |
| 9  | [9,10)   | 660 | 74 | 48 | .1121 | .1447 |
| 10 | [10,11)  | 538 | 53 | 24 | .0985 | .1305 |
| 11 | [11,12)  | 461 | 33 | 22 | .0716 | .1211 |
| 12 | [12,13)  | 406 | 33 | 18 | .0813 | .1113 |
| 13 | [13,14)  | 355 | 29 | 17 | .0817 | .1022 |
| 14 | [14,15)  | 309 | 19 | 23 | .0615 | .0959 |
| 15 | [15,16)  | 267 | 21 | 20 | .0787 | .0884 |
| 16 | [16,17)  | 226 | 17 | 18 | .0752 | .0817 |
| 17 | [17,18)  | 191 | 13 | 12 | .0681 | .0762 |
| 18 | [18,19)  | 166 | 10 | 13 | .0602 | .0716 |
| 19 | [19,20)  | 143 | 11 | 11 | .0769 | .0661 |
| 20 | [20,21)  | 121 | 12 | 11 | .0992 | .0595 |
| 21 | [21,22)  | 98 | 7 | 8 | .0714 | .0553 |
| 22 | [22,23)  | 83 | 2 | 7 | .0241 | .0539 |
| 23 | [23,24)  | 74 | 7 | 8 | .0946 | .0488 |
| 24 | [24,25)  | 59 | 4 | 6 | .0678 | .0455 |
| 25 | [25,26)  | 49 | 2 | 2 | .0408 | .0437 |
| 26 | [26,27)  | 45 | 4 | 2 | .0889 | .0398 |
| 27 | [27,28)  | 39 | 5 | 2 | .1282 | .0347 |
| 28 | [28,29)  | 32 | 2 | 3 | .0625 | .0325 |
| 29 | [29,30)  | 27 | 0 | 1 | .0000 | .0325 |
| 30 | [30,31)  | 26 | 3 | 1 | .1154 | .0288 |
| 31 | [31,32)  | 22 | 1 | 0 | .0455 | .0275 |
| 32 | [32,33)  | 22 | 0 | 2 | .0000 | .0275 |
| 33 | [33,34)  | 20 | 1 | 0 | .0500 | .0261 |
| 34 | [34,35)  | 20 | 0 | 1 | .0000 | .0261 |
| 35 | [35,36)  | 19 | 1 | 5 | .0526 | .0247 |
| 36 | [36,37)  | 13 | 0 | 0 | .0000 | .0247 |
| 37 | [37,38)  | 13 | 0 | 0 | .0000 | .0247 |
| 38 | [38,39)  | 12 | 0 | 1 | .0000 | .0247 |
| 39 | [39,40)  | 11 | 0 | 1 | .0000 | .0247 |
| 40 | [40,41)  | 10 | 0 | 1 | .0000 | .0247 |
| 41 | [41,42)  | 9 | 1 | 0 | .1111 | .0220 |
| 42 | [42,43) | 8 | 0 | 2 | .0000 | .0220 |
| 43 | [43,44) | 6 | 1 | 0 | .1667 | .0183 |
| 44 | [44,45) | 6 | 0 | 1 | .0000 | .0183 |
| 45 | [45,46) | 5 | 1 | 0 | .2000 | .0146 |
| 46 | [46,47) | 4 | 0 | 0 | .0000 | .0146 |
| 47 | [47,48) | 4 | 0 | 0 | .0000 | .0146 |
| 48 | [48,49) | 3 | 1 | 0 | .3333 | .0098 |
| 49 | [49,50) | 3 | 0 | 1 | .0000 | .0098 |
| 50 | [50,51) | 2 | 2 | 0 | 1.000 | .0000 |
| Overall hazard rate | | | | | .2109 | |

Note: Survivor function is calculated over full data and evaluated at indicated times; it is not calculated from aggregates shown at left.

Table 4.
*Modeling results*

| Variables | Model 1 | Model 2 | Model 3 | Model 4 | Model 5 |
|---|---|---|---|---|---|
| *Sponsorship Type* | | | | | |
| Naming Rights | -19.79*** (.01) | -19.79*** (.01) | -19.72*** (.01) | -18.50*** (.02) | -17.53*** (.02) |
| Event Title | -20.52*** (.02) | -20.54*** (.02) | -19.72*** (.02) | -17.64*** (.02) | -15.76*** (.02) |
| League | -18.25*** (.02) | -18.24*** (.02) | -17.96*** (.02) | -17.34*** (.26) | -16.18*** (.02) |
| Olympic Games | -8.05*** (.04) | -8.07*** (.04) | -8.06*** (.04) | -7.86*** (.04) | -7.17*** (.05) |
| World Cup | -9.72*** (.04) | -9.73*** (.04) | -9.76*** (.04) | -9.70*** (.04) | -9.94*** (.04) |
| MLB | 0.96 (.20) | 0.96 (.20) | 0.93 (.20) | 1.03 (.21) | 1.25 (.20) |
| NBA | 0.15 (.15) | 0.15 (.15) | 0.16 (.15) | 0.54 (.17) | 0.92 (.18) |
| NHL | 0.74 (.14) | 0.74 (.14) | 0.72 (.14) | 0.89 (.15) | 1.21 (.17) |
| NFL | -1.54 (.11) | -1.55 (.11) | -1.54 (.11) | -1.35 (.12) | -1.23 (.12) |
| *Economic Conditions* | | | | | |
| Economic Growth (GDP) | | 0.39 (.01) | 0.64 (.01) | 0.62 (.01) | 0.37 (.01) |
| Inflation (CPI) | | 3.42** (.00) | 1.69 (.00) | 1.98* (.00) | 1.88 (.00) |
| *Sponsor Location* | | | | | |
| Africa | | | 2.26* (.29) | 2.52* (.31) | 2.06* (.27) |
| Asia | | | -2.46* (.04) | -2.49* (.04) | -2.43* (.04) |
| Australia | | | 0.97 (.14) | 1.16 (.14) | 0.72 (.13) |
| Europe | | | -0.45 (.03) | -0.43 (.03) | -0.35 (.03) |
| South America | | | 3.69*** (.15) | 3.71*** (.16) | 2.96** (.15) |
| *Sponsor Category* | | | | | |
| Alcoholic Beverage | | | | -2.03* (.06) | -0.81 (.07) |
| Non-Alcoholic Beverage | | | | -2.68** (.08) | -1.57 (.09) |
| Automotive | | | | -3.97*** (.04) | -1.53 (.05) |
| Insurance | | | | -4.81*** (.06) | -4.64*** (.06) |
| Apparel | | | | -3.79*** (.06) | -2.93** (.07) |
| Retail | | | | -2.86** (.05) | -3.10** (.05) |
| Tech | | | | -3.59*** (.04) | -1.36 (.04) |
| QSR | | | | -1.25 (.12) | -0.87 (.14) |
| Food | | | | -2.57 (.06) | -2.18* (.06) |
| Media | | | | 2.28* (.09) | 2.71** (.09) |
| Bank | | | | -5.41*** (.06) | -5.13*** (.06) |
| Credit Card | | | | -2.83** (.10) | -2.00* (.11) |
| Financial Services | | | | -2.14* (.05) | -1.72 (.06) |
| Medical/Hospitals | | | | -1.19 (.14) | -1.36 (.14) |
| Pharmaceutical | | | | 2.15* (.19) | 2.07* (.24) |
| Personal Care | | | | -2.37* (.06) | -1.68 (.06) |
| Airline | | | | -3.08** (.07) | -3.19** (.07) |
| Shipping/Mail | | | | -2.16* (.14) | -1.34 (.16) |
| Utilities/Power | | | | -1.66 (.06) | 0.16 (.06) |
| Hotel | | | | 0.88 (.24) | 1.62 (.23) |
| Betting | | | | -1.91 (.08) | -1.57 (.09) |
| Tire | | | | -0.77 (.07) | 0.92 (.09) |
| Telecom | | | | -2.29* (.06) | -0.69 (.06) |
| *Sponsor Characteristics* | | | | | |
| Regional Proximity | | | | | -4.08*** (.03) |
| Congruence | | | | | -2.23* (.03) |
| Brand Equity | | | | | -4.38*** (.04) |
| B2B | | | | | -2.25* (.03) |
| Publicly-Traded | | | | | -8.02*** (.02) |
| Clutter | | | | | 0.13 (.00) |
| AIC | 77532.2 | 76783.3 | 76777.2 | 76747.2 | 76655.3 |
| BIC | 77604.7 | 76871.9 | 76906.1 | 77061.4 | 77017.9 |
| Log-likelihood | -38382.2 | -38380.6 | -38372.6 | -38334.6 | -38282.7 |
| Wald $\chi^2$ | 1299.45*** | 11.91** | 38.60*** | 105.84*** | 125.87*** |

Results from Cox proportional hazards model. Standardized coefficients are listed, with robust standard errors in parentheses. Standard errors are clustered by sponsorship.
* $p < .05$; ** $p < .01$; *** $p < .001$

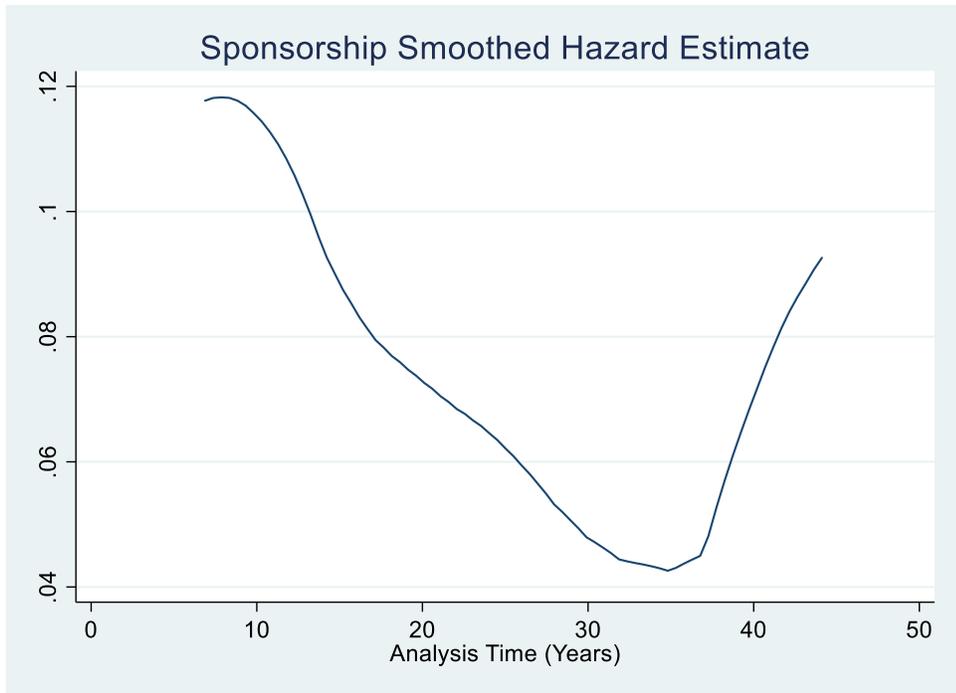

A

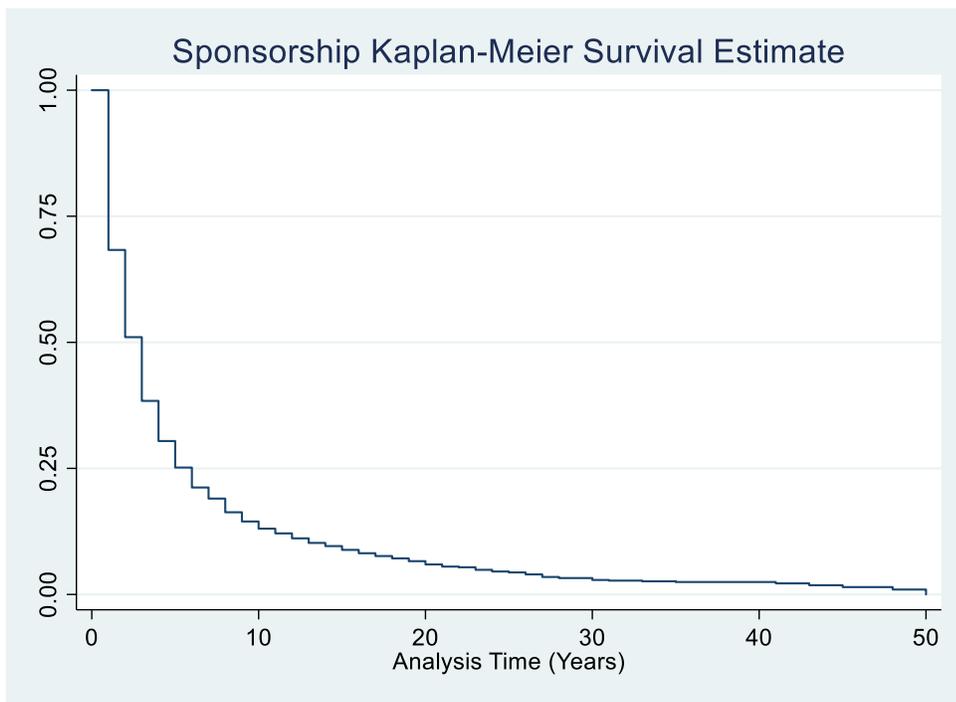

B

*Figure 1.* Graphs of smoothed hazard (Graph A) and survivor functions (Graph B) for sponsorships.

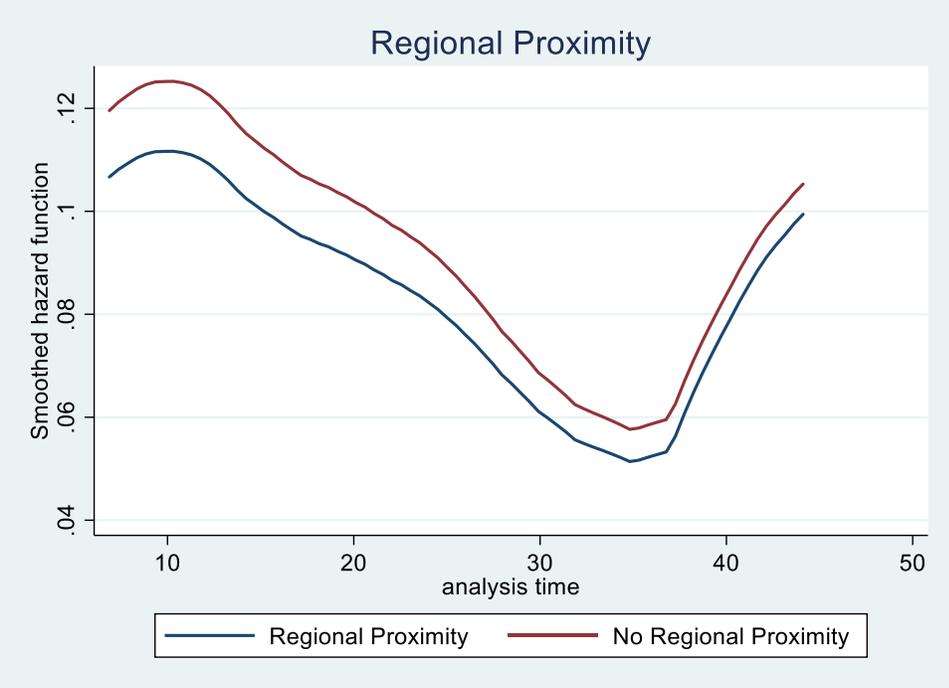

A

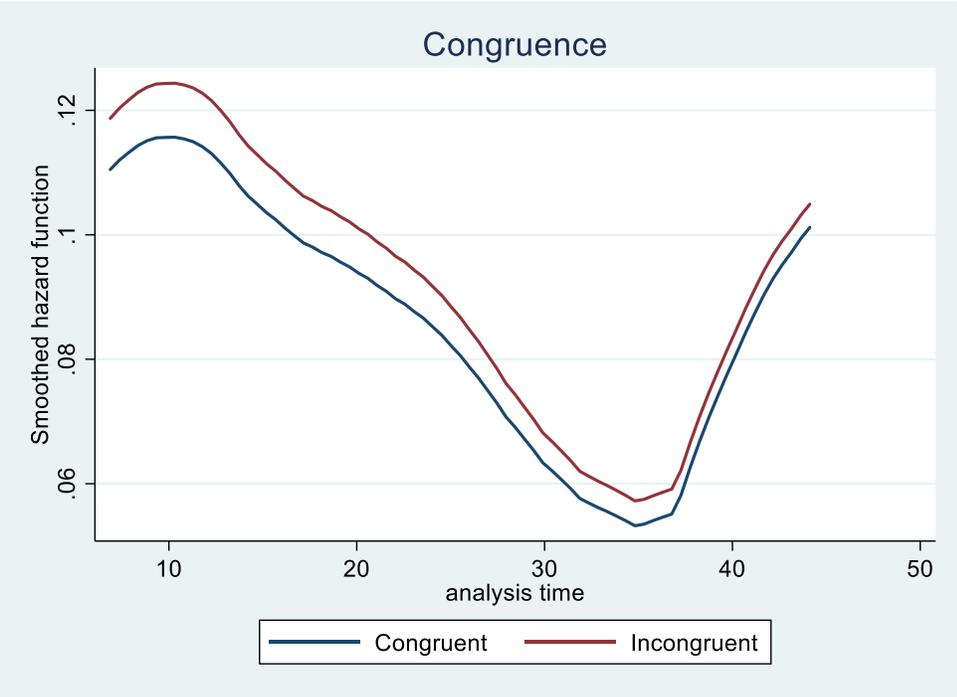

B

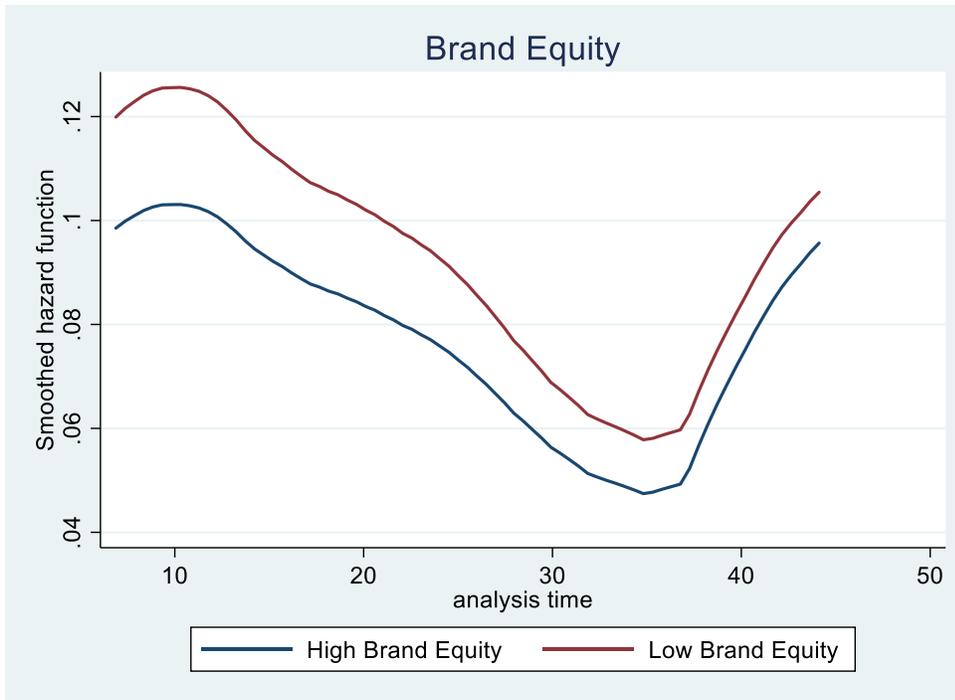

C

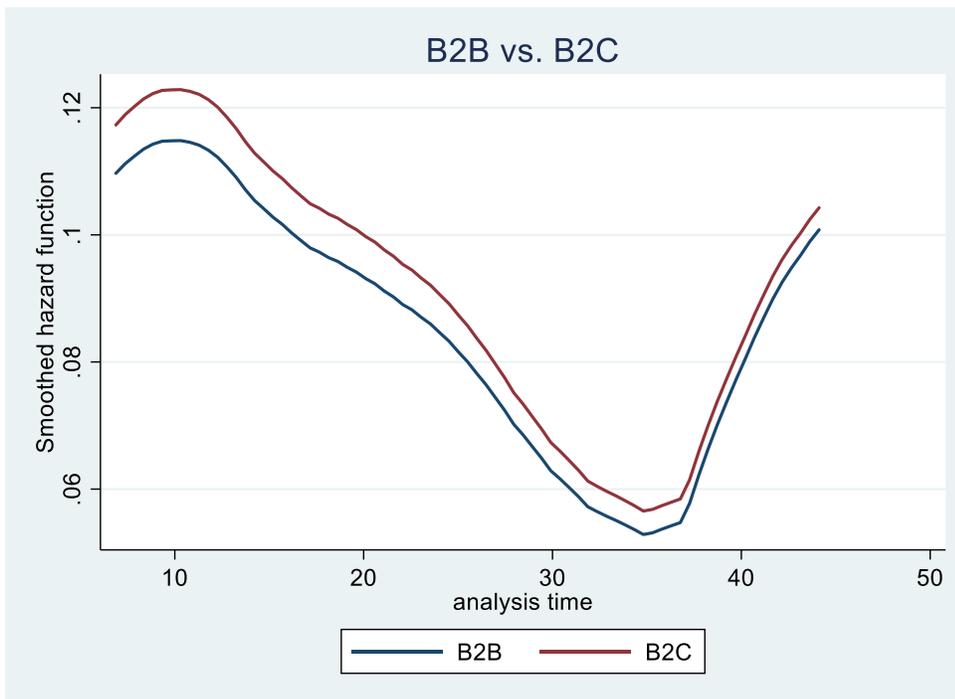

D

*Figure 1.* Graphs of smoothed hazard functions based on sponsor characteristics, including regional proximity (Graph A), congruence (Graph B), brand equity (Graph C), and B2B vs. B2C (Graph D).